\documentclass[aps,prl,showpacs,floatfix,twocolumn]{revtex4}
\usepackage{rotate}
\usepackage{graphicx}
\usepackage{amsfonts}
\newcommand{\km}{k_{max}}
\newcommand{\kmm}{k_{min}}

\newcommand{\kij}{k_ik_j}

\begin{document}

\title{Minimum spanning trees on weighted scale-free networks}
\author{P. J. Macdonald}

\author{E. Almaas}

\author{A.-L. Barab{\'a}si}
\email{alb@nd.edu}

\affiliation{Center for Network Research and Department of Physics,
University of Notre Dame, Notre Dame, Indiana 46556}

\date{\today}

\begin{abstract}
A complete understanding of real networks requires us to understand
the consequences of the uneven interaction strengths between a
system's components.  Here we use the minimum spanning tree (MST) to
explore the effect of weight assignment and network topology on the
organization of complex networks.  We find that if the weight
distribution is correlated with the network topology, the MSTs are
either scale-free or exponential.  In contrast, when the correlations
between weights and topology are absent, the MST degree distribution
is a power-law and independent of the weight distribution.  These
results offer a systematic way to explore the impact of weak links on
the structure and integrity of complex networks.
\end{abstract}


\maketitle

The study of many complex systems have benefited from representing
them as networks \cite{review}, examples including metabolic networks
\cite{jeong00}, describing the reactions in a cell's metabolism; the
protein interaction network \cite{jeong01}, capturing the binding
interactions between a cell's proteins; and the World Wide Web and
email networks \cite{albert99,ebel02} linking web-pages or people
together via URLs or emails.  For these systems there is extensive
empirical evidence indicating that the degree (or connectivity)
distribution of the nodes follows a power-law, strongly influencing
everything from network robustness \cite{robust} to disease spreading
\cite{vespignani01}.  However, to fully characterize these systems, we
need to acknowledge the fact that the links can differ in their
strength and importance
\cite{yook01,goh01,braunstein03,barrat04,toro04}.  Indeed, in a social
network the strength of the relationship between two long-time friends
differs from that between two casual business associates
\cite{granovetter73}; in ecological systems the strength of a
particular pair-interaction between species is crucial for population
dynamics \cite{kilpatrick03}, ecosystem stability \cite{berlow99} and
development in stressed environments \cite{callaway02}.  Thus in most
networks the links are not binary (present or absent), but have a
strength that quantifies the importance of the particular node-to-node
interaction.

The weakest links can carry particular significance in some weighted
networks\cite{granovetter73}.  For instance, the speed of data
transmission between two computers is limited by the link with the
smallest bandwidth (``bottleneck''), or the activity of a metabolic
pathway is determined by the rate of the slowest reaction.
Furthermore, weak links can affect the overall network integrity.  For
example, ecological communities may experience dramatic effects upon
the removal of weak interactors \cite{berlow99}.  To systematically
uncover the location and the role of weak links in a complex network,
we use the minimum spanning tree (MST), which for an $N$ node network
represents the loopless subgraph of $(N-1)$ links that reaches all
nodes while {\em minimizing} the sum of the link weights
\cite{laszlo96,west97,banavar99,banavar00}.  By avoiding the strong
links and preferentially following the weakest ones, the MST selects
the lowest weight backbone of a network.

We start by examining the correlations between weights and network
structure for several real systems, allowing us to construct a model
system whose weight distribution mimics the statistical features of
real networks.  We then show that the large-scale structure of the
MSTs depends on the way the weights are placed in the network: For
systems whose weight distribution is correlated with the network
topology, the MSTs are either scale-free or exponential.  In contrast,
when the correlations between weights and topology are removed, the
MST degree distribution is a power-law with a degree exponent close to
the degree exponent of the original network, independent of the weight
distribution.

\begin{figure}[t]
\centerline{\includegraphics[width=8.6cm]{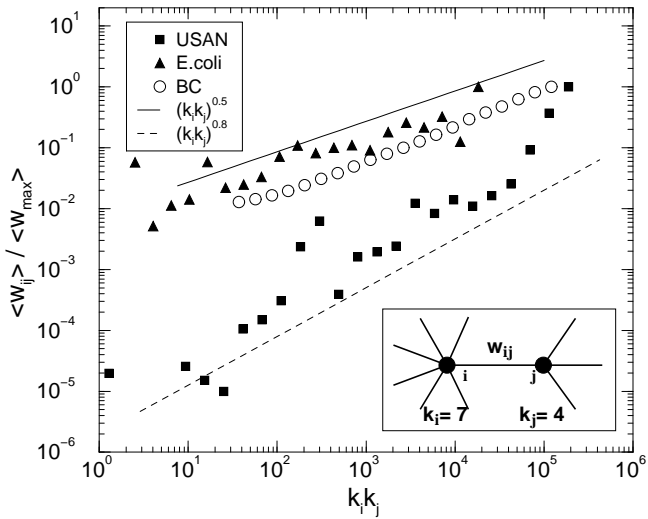}}
\caption{\label{fig:1} The average weight of a link between nodes $i$
and $j$ shown as function of the link end-point degree product $\kij$.
The symbols represent (i) the USAN with the number of passengers as
link weights (filled squares); (ii) the {\it E. coli} metabolic
network with optimized flux as link weight \cite{almaas04} (filled
triangles); (iii) Barab{\'a}si-Albert scale-free model with
betweenness-centrality (BC) as link weights (open circles).  The solid
line ($w\sim (\kij)^{0.5}$) and the dashed line ($w\sim (\kij)^{0.8}$)
serve as guides to the eye.  {\bf Inset:} The weights are determined
by the end-point degrees $k_i$ and $k_j$: (i) $w_{ij} = \kij$, (ii)
$w_{ij} = \max(k_i,k_j)$, (iii) $w_{ij} = \min(k_i,k_j)$ or the
inverse thereof (iv)-(vi).}
\end{figure}

{\it Topology Correlated Weights.}---To uncover the functional
relationship between network topology and link weights, in
Fig. \ref{fig:1} we display the dependence of the weights on the node
degrees for the {\it E. coli} metabolic network, where the link
weights represent the optimal metabolic fluxes \cite{almaas04}; the US
Airport Network (USAN) where the weights reflect the total number of
passengers travelling between two airports between $1992$ and $2002$;
and the link betweenness-centrality (BC), representing the number of
shortest paths along a link for the Barab{\'a}si-Albert (BA)
scale-free model \cite{BA}.  For each of these systems the weight
distributions follow a power-law \cite{goh01} (not shown) and, as Fig
\ref{fig:1} shows, the average link weight scales with the degrees of
the nodes on the two ends of a link as $\langle w_{ij}\rangle \sim
(k_ik_j)^\theta$, similar to the scaling found for the World Airport
Network \cite{barrat04}.

These empirical observations allow us to assign weights to the links
of a network for which we have only the network topology.  To
systematically study the role of the weight distribution on the
structure of the MST we use several weight assignments. (i) First, we
choose $w_{ij} = \kij$ (see inset Fig. \ref{fig:1}).  Note that the
MST generated by this weight assignment is identical to the MST
obtained for weights $w'_{ij} = (w_{ij})^\theta$ with any $\theta >
0$, as it is the rank of the weights and not their absolute value that
determines the MST \cite{dobrin01}.  We have also studied the two
extreme cases of topology-correlated weights, distributed according to
(ii) $w_{ij} \sim \km$ and (iii) $w_{ij} \sim \kmm$, where $\kmm =
\min(k_i,k_j)$ and $\km = \max(k_i,k_j)$ and with $\kmm^2 \leq \kij
\leq \km^2$.  Finally, we investigated the structure of the maximal
spanning trees \cite{kim04} for the above weight choices by
determining the MST after transforming cases (i)-(iii) as $w'_{ij} =
1/w_{ij}$, resulting in the link-weight choices (iv) $w_{ij} \sim
1/\kij$, (v) $w_{ij} \sim 1/\km$ and (vi) $w_{ij} \sim 1/\kmm$.

\begin{figure}[t]
\centerline{\includegraphics[height=7cm]{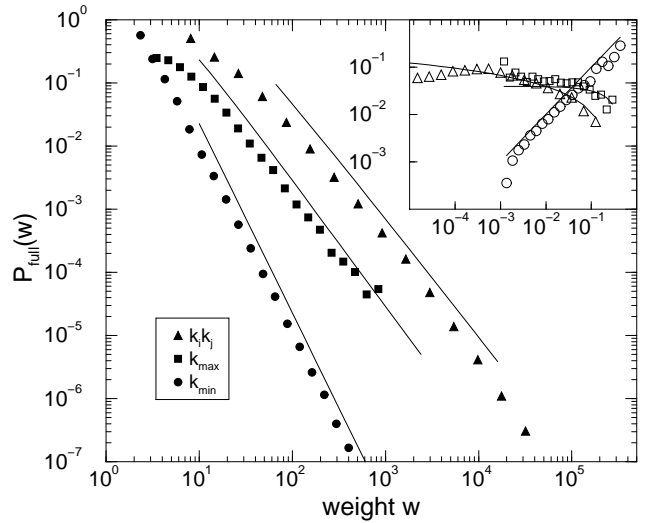}}
\caption{\label{fig:2} Distribution of link weights on $N=10^5$ node
scale-free networks.  Link-weight choice (i) (triangles), (ii)
(squares) and (iii) (circles) are all heavy tailed.  The analytical
predictions (Eqs. (\ref{eq:1}) - (\ref{eq:3})) are indicated as solid
lines. Note that the solid curves have been shifted vertically without
changing the character of the scaling law.  {\bf Inset:} The inverse
weight distributions (iv)-(vi) (triangles, squares and circles
respectively) and the analytical predictions shown as continuous
lines.}
\end{figure}

{\it Weight Distributions.}---To characterize the obtained weighted
networks we first study their weight distribution.  For this, we grow
scale-free networks according to the BA model \cite{BA}, the resulting
networks having a degree distribution $P(k) \sim k^{-\gamma}$ with
$\gamma = 3$.  We then assign a weight to each link according to
(i)-(vi).  For networks whose degrees at the two ends of a link are
uncorrelated we can determine the weight distribution analytically
using order statistics \cite{hogg95}, finding
\begin{eqnarray}
P_{\kij}(w) &=& (\gamma-2)^2~ m^{2(\gamma-2)}~ w^{-\gamma+1}~\ln(w/m^2), \label{eq:1}\\
P_{\km}(w)  &=& 2(\gamma-2) m^{\gamma-2} w^{-\gamma+1} \left[ 1- \left(\frac{w}{m}
                \right)^{-\gamma+2}\right],\label{eq:2}\\
P_{\kmm}(w) &=& 2(\gamma-2) m^{2(\gamma-2)} w^{-2\gamma+3}\label{eq:3}.
\end{eqnarray}
Corresponding expressions for the inverse degree correlations are
obtained after the variable change $w'=1/w$.  For $w_{ij} \sim
(\kij)^\theta$ the relationship between the exponent of the weight
distribution ($P(w) \sim w^{-\sigma}, w\gg 1$), the exponent of the
degree distribution $\gamma$ and $\theta$ is
\begin{equation}
\sigma ~=~ 1 + \frac{\gamma - 2}{\theta},
\end{equation}
valid for cases (i) and (ii).  In Fig.~\ref{fig:2} we compare the
numerically determined weight distributions with the scaling predicted
by our analytical expressions, finding that the numerical curves
display a $w$-dependency close to that of
Eqs. (\ref{eq:1})-(\ref{eq:3}), unaffected by the degree-degree
correlations in the model \cite{krapivsky01,corr}.  Note that
power-law weight distributions like those in Fig. \ref{fig:2} have
been observed for a wide range of network based dynamical processes
\cite{marcio}.

{\it Minimum Spanning Trees.}---The MSTs were generated using Prim's
greedy algorithm \cite{prim57}: starting from a randomly selected
node, at each time step we add the link (and hence a node) with
the smallest weight among the links connected to the already accepted
nodes.  Whenever $m$ links with the same (smallest) weight are
encountered, we break the degeneracy by randomly selecting one among
them with probability $1/m$.

\begin{figure}[t]
\centerline{\includegraphics[height=7cm]{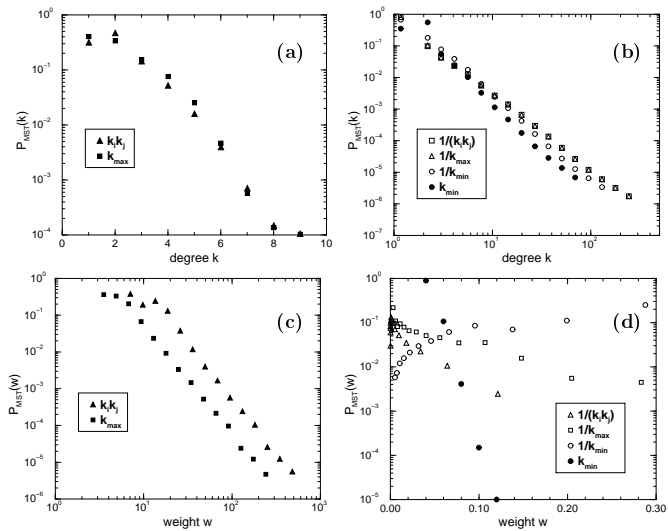}}
\caption{\label{fig:3} Degree and weight distribution of $N=10^4$ node
MSTs. {\bf (a)} The degree distribution for weights proportional to
either $\kij$ (i) or $\km$ (ii) are dominated by an exponential
cut-off, while {\bf (b)} it is heavy-tailed for weights proportional
to $\kmm$ (iii) and inversely proportional to either $\kij$ (iv),
$\km$ (v) and $\kmm$ (vi). {\bf (c)} The distribution of link weights
on the MSTs is a power law for (i), (ii) and (vi) ($w\times 10^3$),
while {\bf (d)} it is dominated by an exponential cut-off for (iii)
($w\times 0.02$), (iv) and (v). For each weight choice we averaged over
$10^4$ different MSTs.}
\end{figure}

\begin{widetext}

\begin{figure}[h]
\centerline{\includegraphics[width=6.9in]{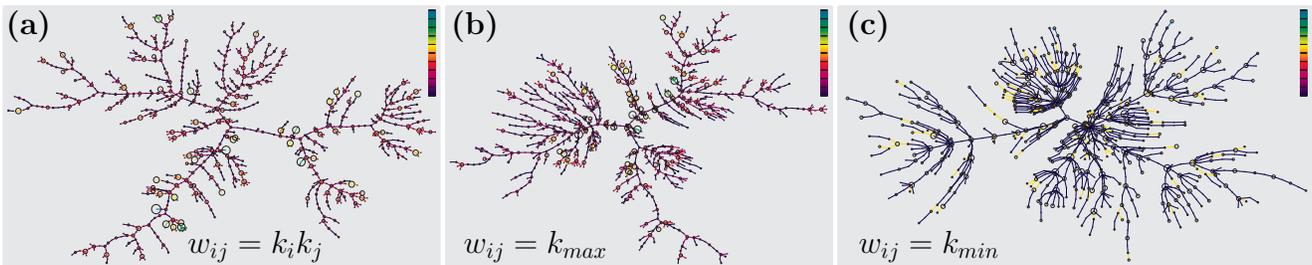}}
\caption{\label{fig:4}(Color online) Minimum spanning trees of a
$N=10^3$ node scale-free network for weight choices {\bf (a)} $\kij$,
{\bf (b)} $\km$ and {\bf (c)} $\kmm$.  The size of a node represents
its degree in the full network, and the color of a link represents its
weight from low (black) to high (green).  Note that the MST degree
distribution is exponential for {\bf (a)} and {\bf (b)} and a power
law for {\bf (c)}.}
\end{figure}

\end{widetext}

The numerical results indicate that the degree distribution of the
resulting MSTs fall into two distinct classes
\cite{kertesz03,kertesz03a}.  Weight choices (i) and (ii) give rise to
exponential MST degree distributions (Fig. \ref{fig:3}a), while
choices (iii)-(vi) result in power-law distributed MST degrees
(Fig. \ref{fig:3}b).  We can understand the exponential nature of the
(i) and (ii) MSTs through the following argument: Since the MST tends
to avoid links with large weights, it effectively shuns the hubs for
the cases $w_{ij}=\kij$ and $w_{ij}=\km$, utilizing instead, whenever
possible, links connecting low degree nodes (Fig. \ref{fig:3}a).
Consequently, all the hubs are marginalized and the MST degree
distribution must have a narrow range.  This argument is supported by
Fig. \ref{fig:4}a and b, where we show examples of MSTs for weight
choices (i) and (ii) respectively.  The sizes of the nodes in the
figure reflect their degree in the original network.  It is evident
that the majority of the hubs are located on the branches ($k=1$
degree nodes) of the MST (Fig. \ref{fig:4}).  This reliance on the
small nodes and tendency to avoid the hubs forces the MSTs generated
by method (i) and (ii) to be very similar to each other.  Indeed, we
find that for a given network but weights created by methods (i) and
(ii), 87\% of the links in the two MSTs are in common.  This explains
the similar visual appearance of the two MSTs (Fig. \ref{fig:4}a and
b).

The second class of MSTs is well represented by weight choices
(iii)-(vi), resulting in power-law MST degree distributions.  The
similarity between weight schemes (iv)-(vi) is emphasized by the fact
that their MST degree distributions follow a power-law with the same
exponent $\gamma = 2.4$ \cite{kim04} (Fig.~\ref{fig:3}b).  Indeed, the
links with the lowest weights are now connected to the hubs of the
original network, and the MST grows utilizing these hubs extensively.
Hence, the hubs of the full network experience only a slight reduction
in their degree and are found at the center of the resulting MSTs
(Fig. \ref{fig:4}c), while the intermediate-degree nodes sustain large
losses of neighbors and are found at the surface of the network with
one or two neighbors (Fig. \ref{fig:4}c).

The distribution of link weights on the MST also displays two distinct
behaviors, being either power-law (Fig. \ref{fig:3}c) or exponential
(Fig. \ref{fig:3}d).  It is interesting to note that MSTs with
exponential degree distribution (Fig. \ref{fig:3}a) display power-law
weight distributions with exponents $\sigma = 3.1$ (case (i)) and
$\sigma = 3.0$ (case (ii)) (Fig. \ref{fig:3}c).  On the other hand,
for weight choices (iii)-(v) the degree distribution of the MST is
power law and the MST weight distribution is exponential or stretched
exponential (Fig. \ref{fig:3}d).  For (vi) $w_{ij} = 1/\kmm$ both the
degree and the weight distribution of the MST are scale free.
Finally, if the link weights are distributed uniformly and randomly
the resulting MSTs have a power law degree distribution and an
exponentially tempered weight distribution \cite{kertesz03a}.

In order to investigate the effect of the degree correlations on the
MSTs for weight choices (i)-(vi), we randomized the weights of the
original network by randomly selecting pairs of links and exchanging
their weights until all correlations between weights and the local
network
topology were lost.  Invariably, the resulting MSTs were
scale-free with a degree exponent similar to that of the original
network, $\gamma \approx 3$.  The local structure of the MSTs 
are very different, however, with only $52$\% of the links staying the
same in a pairwise comparison between MSTs with weight choices (i) and
(ii), suggesting that the functional form of the weight distribution
is inconsequential for the degree distribution of the MSTs.  To
understand this we recall that only the {\em ranking} of the link
weights and not their absolute value matters \cite{dobrin01}.
Therefore, by removing the correlations between the local network
structure and weights we effectively map the problem onto that of
weights being uniformly random.  Indeed, the degree distribution of
the MST in this case is also power-law with $\gamma \approx 3$
\cite{kertesz03a}.  However, the MST weight distributions continue to
depend on the weight distribution of the original network.

{\it Discussion.}---As networks play an increasing role in the
exploration of complex systems, there is an imminent need to
understand the interplay between network dynamics and topology.  While
focusing on the MSTs of scale-free networks, our results emphasize the
significance of correlations between link-weights and local network
structure.  We find that if correlations are present, two classes of
MSTs exist, following either a power-law or an exponential degree
distribution.  The removal of correlations renders the MSTs
scale-free, independent of the choice of the weight distribution.
This result raises interesting questions regarding our ability to
quantify the influence of weights.

Our findings could serve as a natural starting point towards the
systematic exploration of weighted networks.  For example, while we
have assumed that the weights are static, incorporating their
time-dependence may reveal novel dynamical rules.  Second, we model
the weights as solely dependent on the topology, potentially
overlooking correlations among the weights themselves.  Uncovering the
role of such correlations remains a challenge for future research.

We thank J. Kert{\'e}sz, P. L. Krapivsky and S. Havlin for
discussions.  We also thank M. A. de Menezes for sharing the US
airport data.  This work has been supported by grants from DoE (E.A.)
and the R.E.U. program at Notre Dame (P.J.M).


\begin{thebibliography}{100}
\bibitem{corr} Taking into account the degree-degree correlations of
our model \cite{krapivsky01}, we find the same $w \gg 1$ limiting
behavior as that of Eqs. (\ref{eq:1})-(\ref{eq:3}).
\end{thebibliography}
\end{document}